\documentclass[%
 reprint,
 amsmath,amssymb,
 aps,
 showkeys,
]{revtex4-2}

\usepackage{graphicx}
\usepackage{dcolumn}
\usepackage{bm}
\usepackage{hyperref}

\begin{document}

\title{\textit{Paperfetcher}: A tool to automate handsearch for systematic reviews}

\author{Akash Pallath}
\thanks{Authors are listed alphabetically. Both authors contributed equally to this work.}
\affiliation{School of Engineering and Applied Science, University of Pennsylvania, Philadelphia, PA - 19104\\\rm\url{apallath@seas.upenn.edu}}

\author{Qiyang Zhang}
\thanks{Authors are listed alphabetically. Both authors contributed equally to this work.}
\affiliation{School of Education, Johns Hopkins University, Baltimore, MD - 21218\\\rm\url{qzhang74@jhu.edu}}

\begin{abstract}
Handsearch is an important technique that contributes to thorough literature search in systematic reviews. Traditional handsearch requires reviewers to systematically browse through each issue of a curated list of field-specific journals and conference proceedings to find articles relevant to their review. This manual process is not only time-consuming, laborious, costly, and error-prone, but it also lacks replicability and cross-checking mechanisms. In an attempt to solve these problems, this paper presents a free and open-source Python package and an accompanying web-app, \textit{Paperfetcher}, to automate handsearch for systematic reviews. With \textit{Paperfetcher}'s assistance, researchers can retrieve articles from designated journals within a specified time frame with just a few clicks. In addition to handsearch, this tool also incorporates snowballing in both directions. \textit{Paperfetcher} allows researchers to download retrieved studies as a list of DOIs or as an RIS database to facilitate seamless import into citation management and systematic review screening software. To our knowledge, \textit{Paperfetcher} is the first tool that automates handsearch with high usability and a multi-disciplinary focus.
\end{abstract}


\maketitle


\section{Introduction}
Literature search is an early and crucial step for performing a comprehensive systematic review \cite{cooper2018, armstrong2005, richards2008}. The process of literature search often begins with a field-related bibliographic database search, which retrieves studies from electronic records that index journals and non-journal sources \cite{kugley2017}. Limiting literature search to bibliographic search risks missing high quality papers \cite{richards2008, papaioannou2010} that were not indexed with identifiable terms \cite{dickersin1994}, formatted as abstracts or letters \cite{armstrong2005, richards2008}, located in supplement editions of journals \cite{armstrong2005}, or not included in electronic databases \cite{blumle2008}. Handsearch is an additional step that reviewers often undertake to identify such studies, which involves systematically browsing through the tables of contents from a curated list of field-specific journals, abstracts, and conference proceedings in order to gather papers relevant to the synthesis topic \cite{kugley2017}. To ensure comprehensive information retrieval and transparent reporting, it is important to conduct literature search in a systematic and exhaustive manner \cite{cochrane2008}. Combining handsearch with bibliographic database search decreases the likelihood of missing major relevant studies and therefore underpins a solid foundation for the systematic review \cite{hopewell2007}.

Despite its importance, the prevailing practice of handsearch awaits urgent development. Advancing handsearch strategy is also timely due to the rapid increase in studies available online and the subsequent increase in the amount of time required to retrieve papers. Traditionally, handsearch requires reviewers or hired volunteers to manually read through the tables of contents and abstracts of tens to hundreds of journal issues, supplementary materials, and conference proceedings \cite{kugley2017, cochrane2008}. Past literature agrees on the time and labor-intensive nature of handsearch, reporting time spent on this task ranging from an hour per journal volume \cite{moher1995} to 185 hours spent for 10 journals \cite{glanville2012}. This procedure is not only time-consuming, laborious \cite{kugley2017}, and costly \cite{moher1995}, but more importantly, it is error-prone due to human fatigue \cite{adams1994}, lacks replicability and an easy cross-checking mechanism due to its cumbersome manual nature.

To address these problems, we developed a freely available Python package and an accompanying web-app – \textit{Paperfetcher} – that automates handsearch to increase efficiency and ensure replicability. \textit{Paperfetcher} automatically fetches works from a user-selected list of journals within a user-specified timeframe. \textit{Paperfetcher} not only returns article metadata, but it also generates a report of the parameters used to perform the handsearch, which can assist with replicability. Automation of handsearch facilitates less error-prone systematic review, and more importantly, enables researchers to focus their energy on the screening process rather than the search process. \textit{Paperfetcher} addresses the urgent need for a more replicable, robust, and time-efficient method to conduct handsearch.

In addition to handsearch, \textit{Paperfetcher} also includes a snowballing function for forward and backward citation chasing \cite{cooper2017, lefebvre2019, jalali2012, greenhalgh2005}. Forward citation chasing searches for articles that cite a given article of interest while backward citation chasing searches for articles cited by the article of interest. Snowballing, which is an umbrella term for citation chasing in both directions, is a very useful supplementary search strategy \cite{papaioannou2010}. Researchers have found that it identifies 51\% of studies in systematic reviews \cite{greenhalgh2005}. We implemented snowballing in \textit{Paperfetcher} due to its important contribution to literature search.

While there are several existing tools to automate snowballing (further discussed in Section \ref{sec:discussion}), to the best of our knowledge, there is no other tool in the current market that provides an easy-to-use interface to automate handsearch.

\section{How \textit{P\lowercase{aperfetcher}} works}
\textit{Paperfetcher} consists of a) a Python package which implements the handsearch and snowballing algorithms to retrieve raw data and convert it to various output formats (such as text and Research Information Systems (RIS) format), and b) a web-app that provides an easy to use graphical interface for the package. In the following subsections, we describe how the Python package and the web app work.

\begin{figure*}
    \centering
    \includegraphics[width=0.6\textwidth]{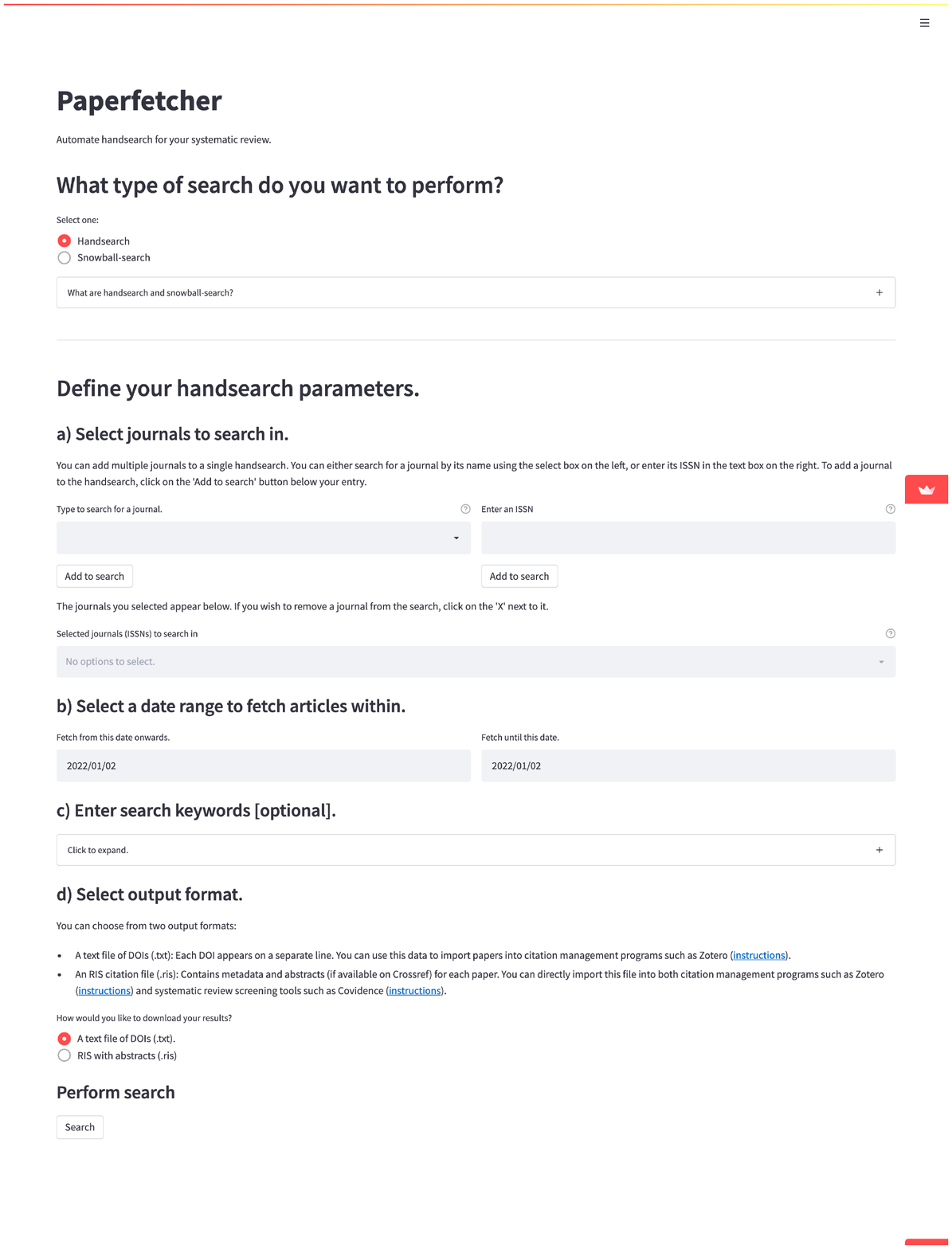}
    \caption{Handsearch user-interface in the \textit{Paperfetcher} web-app.}
    \label{fig:handsearch}
\end{figure*}

\begin{figure*}
    \centering
    \includegraphics[width=0.6\textwidth]{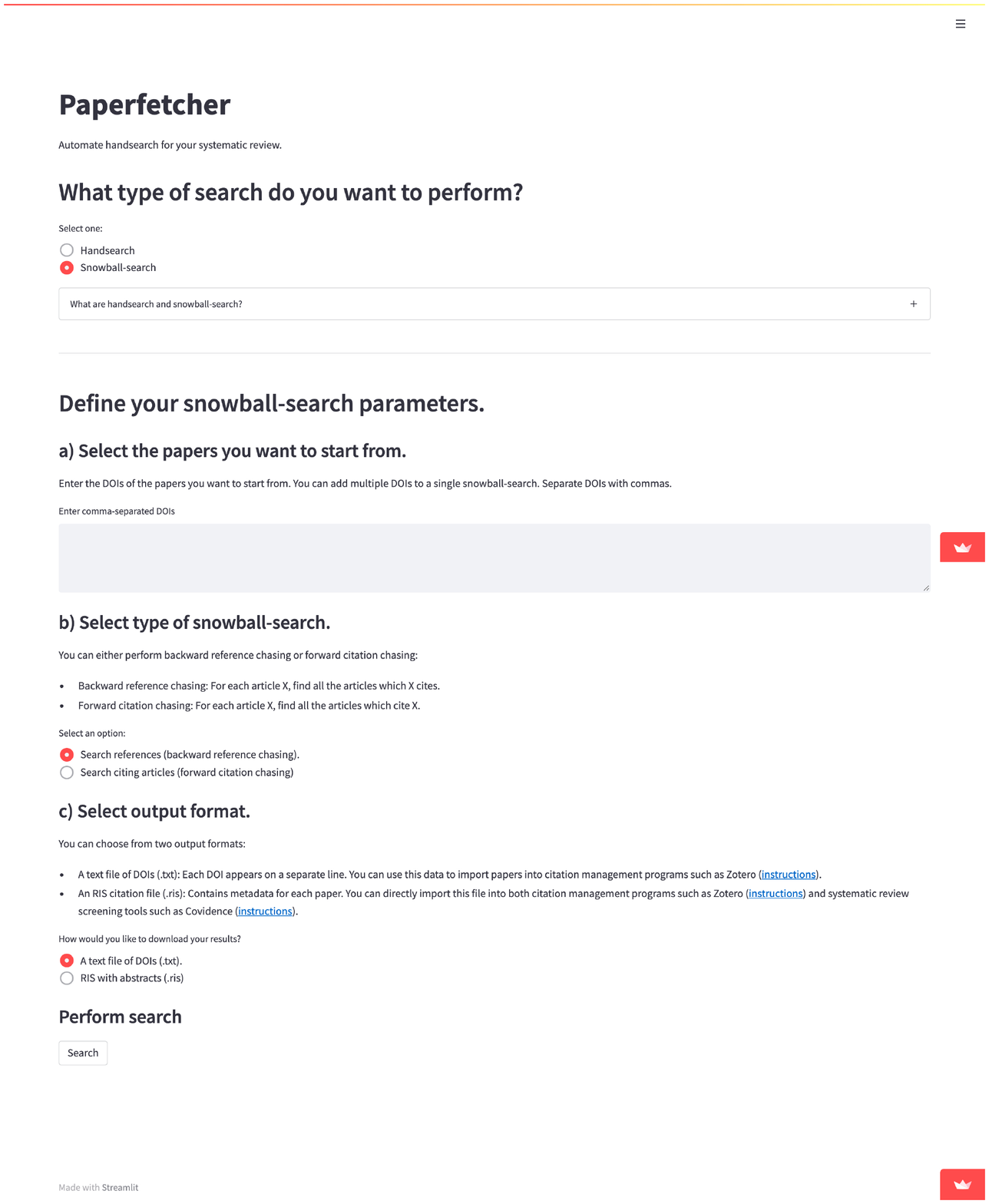}
    \caption{Snowballing user-interface in the \textit{Paperfetcher} web-app.}
    \label{fig:snowballing}
\end{figure*}

\subsection{Data sources}
\textit{Paperfetcher}'s handsearch algorithm queries the database of academic content registered with Crossref. Crossref is an official digital object identifier (DOI) registration agency of the International DOI Foundation \cite{howells2006}. Research suggests that Crossref is the most robust and holistic implementation of the DOI model \cite{howells2006}, and covers a wide range of disciplines. Among the 11 official DOI registration agencies, only Crossref, DataCite, and Multilingual European DOI Registration Agency (mEDRA) cover English materials related to scholarly and professional research content \cite{doi2021}. We excluded DataCite as it only indexes research data and not research articles or reports \cite{doi2021}. As mutual collaboration between mEDRA and Crossref allows DOIs registered with mEDRA to be deposited on the Crossref platform \cite{medra2021}, we excluded mEDRA to avoid redundancy. Since \textit{Paperfetcher} is primarily designed for English-reading researchers, we excluded DOI agencies that contain materials of other non-English languages, such as China National Knowledge Infrastructure \cite{cnki2021}, or which index non-academic content, such as the Entertainment Identifier Registry \cite{eidr2021}. Other non-English DOI agencies can be added in future versions of \textit{Paperfetcher} once demand for multilingual research content arises.

Statistics updated on Dec 31 2021 show that Crossref covers more than 1.3 million published and unpublished content types \cite{crossref2021}, such as journals, books, conference proceedings, dissertations, working papers, technical reports, and data sets. It provides more than 90,000 records of journals and more than 80,000 records of conference proceedings to empower an effective handsearch \cite{crossref2021}. A comparison of different databases, including Scopus, Web of Science, Dimensions, Crossref, and Microsoft Academic, concluded that Crossref is a bibliographic data source that is of significant interest for bibliometric analyses and is becoming increasingly valuable over the years \cite{visser2021}. As compared to the widely-used Scopus database, Crossref covers a larger number of documents that have been published in journals \cite{visser2021}.

\textit{Paperfetcher}'s snowballing algorithm queries both the Crossref database and COCI, the OpenCitations Index of Crossref open DOI-to-DOI citations. COCI is a database derived from Crossref, which contains more than 1.2 billion DOI-to-DOI citation pairs of Crossref-deposited articles with open citations \cite{opencitations2021, heibi2019}. The COCI database is updated periodically to add more citations. As of January 2, 2022, the last update to the database was on November 25, 2021 \cite{opencitations2021}.

\subsection{Handsearch}
Each handsearch has three inputs: the International Standard Serial Number (ISSN) of the journal to fetch articles from, a date range within which to fetch articles, and an optional list of keywords to refine the search. We do not recommend using keywords when performing handsearch for systematic reviews as there is a possibility of missing relevant studies which were not indexed in Crossref with matching keywords. However, this input can still be useful for performing literature searches for specific purposes, or if the reviewers are time-constrained and have to filter through a large number of matching articles. 

\textit{Paperfetcher} queries the Crossref database through its Representational State Transfer Application Programming Interface (REST API) with these three inputs to retrieve metadata – containing fields such as article title, authors, journal, publisher, publishing date, abstract, keywords, etc. – of all articles from the journal with the specified ISSN, within the selected date range, and matching the list of keywords (if specified). The \textit{Paperfetcher} web app can iterate through a list of journals, perform handsearch for each of them, and merge the retrieved metadata into a single dataset of search results.

\subsection{Snowballing}
\textit{Paperfetcher} requires a list of DOIs as input for both forward and backward snowballing. For backward snowballing, it queries the Crossref REST API to retrieve DOIs of papers cited by the input DOIs. The Crossref REST API only returns information about articles cited by a given paper, and not articles citing it. The COCI REST API, however, can return this data, therefore \textit{Paperfetcher} uses COCI for forward snowballing. In both cases, \textit{Paperfetcher} compiles the union of the retrieved DOIs into a dataset of search results.

\subsection{Data export}
The \textit{Paperfetcher} Python package can convert the search results into several formats, such as a text file of DOIs, a CSV file or Excel spreadsheet with rows containing article title, authors, journal, DOI, URL, etc. for each article, RIS databases, and even pandas DataFrames \cite{mckinney-proc-scipy-2010, reback2020pandas} for further data analysis and processing in Python or R. The \textit{Paperfetcher} web-app implements two of these options – it can either export results to an RIS file or to a text file of DOIs. \textit{Paperfetcher} uses Crossref's content negotiation service \cite{contentnegotiation} to retrieve metadata in the RIS format.

\begin{figure*}
    \centering
    \includegraphics[width=0.5\textwidth]{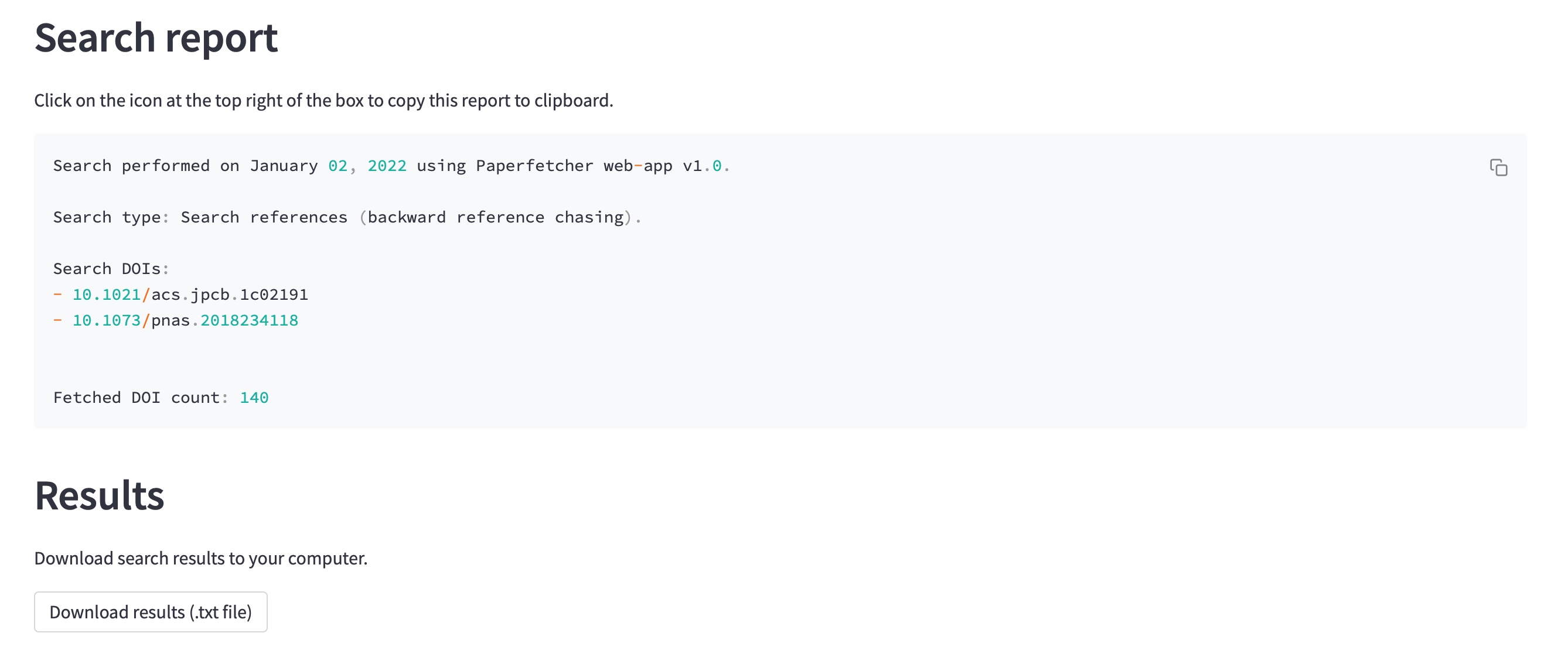}
    \caption{Search results and report in the \textit{Paperfetcher} web-app.}
    \label{fig:results}
\end{figure*}

\section{Using \textit{P\lowercase{aperfetcher}}}

\subsection{Using the web-app online}

\textit{Paperfetcher}'s web-app, based on the Streamlit app framework, is available online at \url{https://share.streamlit.io/paperfetcher/paperfetcher-web-app/main/paperfetcher_app.py}. The web-app can also be run offline, as described in Section \ref{sec:using-offline}. The highlight of this web app is that it has an easy-to-use graphical interface and requires no programming experience. To perform handsearch (Figure \ref{fig:handsearch}), users just need to fill out a few search parameters –  journal names or journal ISSNs, a date range to fetch articles within, optional search keywords, output format – and click on the ‘Search' button. To perform snowballing (Figure \ref{fig:snowballing}), users only need to enter comma-separated DOIs of selected papers, select the type of snowballing (forward or backward), select output format, and click on the ‘Search' button.

\subsection{Exporting data}
After performing a search, \textit{Paperfetcher} gives users the option to download their data either as a text file of DOIs or as an RIS file, based on the output format selection they made (Figure \ref{fig:results}).

RIS is a standardized format that allows data exchange among different citation management software \cite{usaid2018}. RIS data exported from \textit{Paperfetcher} can be seamlessly imported into citation management software such as EndNote, Mendeley, Paperpile and Zotero, and into Systematic Review Screening tools such as ASReview \cite{vandeschoot2021}, Covidence, DistillerSR, and EPPI-Reviewer \cite{eppi2020} (for a comprehensive review of screening tools, see Zhang and Neitzel \cite{zhang2021}). DOIs from exported text data can be bulk-imported into citation management tools such as Zotero in order to fetch missing abstracts or other important metadata.

\subsection{Reporting handsearch parameters}
\textit{Paperfetcher} generates a report of the parameters used to perform the search and the number of papers fetched, which is displayed to users after performing the search. Figure \ref{fig:results} shows a sample report for snowballing. These reports provide a mechanism for reviewers to document their search for cross-checking and reproducibility.

\subsection{Using the web-app offline}
\label{sec:using-offline}
Privacy-conscious users may opt to use the \textit{Paperfetcher} web-app offline. To do so, they must first install and then run the app from source. The source code and installation instructions for the app are available online at \url{https://github.com/paperfetcher/paperfetcher-web-app}.

\subsection{Using the Python package}
Python programmers might prefer to directly use \textit{Paperfetcher}'s Python package as it can export search results to more data formats than the app. Users who wish to set up data pipelines or perform additional data analysis in Python or R will find the package's ability to export data to pandas DataFrames particularly useful. Such users can install the package directly from the Python Package Index (\url{https://pypi.org/project/paperfetcher}). Developers who are interested in modifying \textit{Paperfetcher}'s source code can clone the source repository from GitHub at \url{https://github.com/paperfetcher/paperfetcher}. Documentation for the Python package can be found at \url{https://paperfetcher.github.io/paperfetcher}.

\section{Discussion}
\label{sec:discussion}
There are a few existing tools that automate snowballing. Some tools only enable uni-directional citation chasing. For example, Sci-Finder supports backward reference chasing, but it is not free of charge and only focuses on literature related to chemicals, drugs, and substances \cite{vogel2021}. Some databases, including Web of Science, Scopus, and Google Scholar, enable forward citation chasing. We only found two tools that support citation chasing in both directions: SpiderCite and citationchaser. SpiderCite is part of the free online suite of tools SR-Accelerator, and allows users to import and export references in EndNote XML, RIS, and BibTeX formats \cite{SRaccelerator2021}. Citationchaser is an R package that is free of charge, open source, and provides an easy-to-use Shiny app interface, where users can input DOIs of articles to search and export data in RIS format \cite{haddaway2021}. However, to date, all the above tools lack functions to automate handsearch. \textit{Paperfetcher} serves to complement these tools to help reviewers identify as many relevant studies as possible.

Although \textit{Paperfetcher} has the potential to make significant contributions to the field of systematic review, the current version is limited in a few ways. First, there can be a lag between when a paper is first published online and when its DOI is deposited in Crossref. In such a case, \textit{Paperfetcher} will fail to find the paper. Second, certain publishers do not make their abstracts and references publicly available on Crossref. In the former case, reviewers can use citation management software to retrieve missing abstracts. The latter case, however, is more problematic as this will result in relevant studies being missed by \textit{Paperfetcher}'s snowballing function \cite{visser2021}. In addition, as updates to COCI lag behind updates to Crossref, \textit{Paperfetcher}'s forward citation chasing function might miss newly published citing articles. Citationchaser, which queries the Lens.org database (consisting of articles from PubMed, PubMed Central, CrossRef, Microsoft Academic Graph and CORE), has access to more reference information. As the Lens.org API requires a subscription fee, and \textit{Paperfetcher} was developed without any funding, we had no choice but to exclude paid APIs. In the future, should we have access to funding or resources, we can incorporate additional APIs (such as Lens.org or Web of Science) to improve \textit{Paperfetcher}'s snowballing function.

In a nutshell, \textit{Paperfetcher} provides a free, easy-to-use, and efficient method to automate handsearch. We believe that it can significantly reduce the time reviewers spend on literature search and can improve transparency in synthesis reporting across disciplines.

\nocite{*}

\bibliography{ref.bib}

\end{document}